\documentclass[a4paper,10pt]{article}
\usepackage[utf8x]{inputenc}
\usepackage[english]{babel}
\usepackage{graphicx,amssymb,amsmath,amsthm,amscd,stmaryrd}
\usepackage{epsfig,wrapfig,float}
\usepackage{natbib}
\usepackage{multicol,setspace}

\bibpunct{(}{)}{;}{a}{,}{,}

\makeatletter
 \renewcommand{\section}{\@startsection{section}{1}{\z@}{-3.5ex \@plus -1ex \@minus -.2ex}{2.3ex \@plus.2ex}{\normalfont\normalsize\bfseries}}
 \renewcommand{\subsection}{\@startsection{section}{1}{\z@}{-3.5ex \@plus -1ex \@minus -.2ex}{2.3ex \@plus.2ex}{\normalfont\small\bfseries}}
\makeatother

\usepackage{caption}
\newcounter{abcd}[footnote]

\addto{\captionsenglish}{%
 }

\begin{document}

\begin{center}

\begin{doublespace}
\begin{Large} \textbf{Reconstruction of the 3D structure and developmental history of plant cells and tissues}
\end{Large}           
\end{doublespace}

\vspace{0.5cm}
\setcounter{abcd}{0}
Ivan V. Rudskiy\footnote{Corresponding author: Ivan V. Rudskiy, Institut des Hautes \'Etudes Scientifiques, Le Bois-Marie, 35, route de Chartres, 91440
Bures-sur-Yvette, France. E-mail: rudskiy@ihes.fr, ivrudskiy@mail.ru}$^{1,2}$, 
Nadejda V. Khodorova$^{1,3}$

\vspace{0.2cm}
\begin{footnotesize}

\noindent$^{1}$ Komarov Botanical Institute of the Russian Academy of Sciences, St-Petersburg, Russia

\noindent$^{2}$ Institut des Hautes \'Etudes Scientifiques, Bures-sur-Yvette, France

\noindent$^{3}$ Jules Verne University of Picardie, Amiens, France
 
\end{footnotesize}

\end{center}

\begin{abstract}

Modern spatial microscopy has provided developmental biology with powerful research tools. However, the recent significant technological breakthroughs have inevitably led to technical “bottlenecks”
for the application of the new procedures to other tasks. We have developed a protocol for the 3D reconstruction of plant organelles and tissues from stacks of two-dimensional images obtained by means
of a wide range of electron transmission and light microscopy. This protocol can be applied to a large set of primary data: live, fixed or even paleobotanical material. For tissue reconstruction, we
have also developed a protocol of cell lineage tracing based on the geometrical properties of plant cells. For the tissue sample consisting of $n$ cells, this approach allows the reconstruction of up
to $n-1$ cell divisions from the previous cell generations. Our protocol complements the limitations of clonal analysis and recent real-time approaches of lineage tracing. It works well in the case
of tissues with directed and non-directed growth. Using these protocols, the subcellular relations between the complex surface of plastids with stromules and mitochondria in the tuber parenchyma cells
in \textit{Corydalis bracteata} have been revealed. Moreover, for the embryo ($n=17$) and the seedling shoot apex ($n=208$) of \textit{Calla palustris}, the cellular architecture and full cell
genealogy of up to 15 cell generations have been reconstructed. Three relatively independent cell lineages with stem cells located at the surface of the shoot apex have been found. The possibility of
using a wide range of software, including open-source projects, is discussed. 

\end{abstract}

\vspace{0.1cm}
\begin{small}

\vspace{0.5cm}

\begin{multicols}{2}

\section*{INTRODUCTION}

The spatial organization of biological objects plays a central role in many modern structural research projects. 3D reconstruction has found suitable applications at different organizational levels of
plants and animals, from organelles to tissues and organ systems. The improvement of methods in this sphere has resulted in the creation of new complex techniques, like confocal microscopy, light
sheet microscopy and electron tomography, and special software for data analysis and interpretation. The complexity of analytical algorithms has led to the automation of data collection and processing
and finally to simplifying the user’s work. This situation is very useful for an extensive range of routine scientific work and for express diagnostics in medicine (John and McCloy 2004). In
developmental biology, the real-time acquisition and analysis of morphological data obtained from the living object is of great importance enabling one to trace all developmental cell movements (Heid
et al. 2002) as well as tracing the cell lineages of some model animals in the course of embryonic development up to millions of cells (Keller et al. 2008). The application of sophisticated
mathematical algorithms is the only way to represent 4D or even 5D spatial anatomical, temporal and physiological data (Rosset et al. 2006) and to achieve an optical resolution beyond the diffraction
limits, up to 100-150 nm like in 4-Pi technology (Bewersdorf et al. 2004) or to enhance the resolution range in electron tomography up to 2-3 nm (Frank et al. 2002).

Indeed, modern biological microscopy, especially the routine type, tends to be a “virtual” microscopy with a high dependence on high cost devices and software. The significant technological
breakthroughs in microscopy inevitably lead to new technical “bottlenecks” or limitations for the application of the same procedures to other objects and for the resolving of other scientific
questions (Truong and Supatto 2011). In the present paper, we provide some new possibilities of 3D technologies applicable to “virtual” microscopy as well as to the traditional “two-dimensional
microscopy” like transmission electron microscopy (TEM) and light transmitted microscopy (LTM) which are widely used in modern structural botany. Our aim is to resolve two issues: (1) to reveal the
detailed spatial shape properties of plastids with the help of TEM and of cells in plant tissue with the help of LTM, and (2) to reconstruct the genealogical relations between all the cells in a plant
tissue sample.

The questions we raise here are topical but have been only partially addressed in recent investigations. Contouring or image segmentation is crucial for the proper definition of the spatial structure
of the object of interest in microscopic data. There are several basic approaches of automatic segmentation such as thresholding, watershed transform and region growing. However, the universal
solution of this problem is not feasible and the human visual detecting system is still the most reliable (Khairy and Keller 2011). After investigating a wide range of program packages dedicated to
the conventional 2D and 3D computer raster and vector graphics as well as microscopy-oriented software (Table 1), we have developed a universal protocol oriented to manual work in general, but which
is independent of the device and the software used, for the reconstruction of plant organelles, and the spatial organization of cells and tissues from the stacks of their two-dimensional images. This
approach enables us to obtain not statistical but rather topological information about plant cells and organelles, like the local neighborhood of structures and interruption of surface smoothness.
These very properties, known as positional information, are essential in the description and explanation of the morphogenetic processes. It has been recently shown that the local cell-to-cell
interactions (and undoubtedly interactions between organelles) are even more important for specifying cell fate than overall tissue and organ polarity in plants (Sauer et al. 2006) as well as in
animals (Bischoff and Schnabel 2006).

Higher plants are unique organisms which, in most cases, enable us to trace back all the previous divisions of tissue cells according to the cell wall spatial position (Barlow et al. 2001).
Unfortunately, to date, there are no available protocols or software ready to be applied to the revelation of the developmental history of tissue samples based on plant cell geometry. Only real-time
observation techniques are capable of providing fragmentary information about cell divisions occurring during the observation period. Among these approaches are shoot apex surface replicas (Dumais and
Kwiatkowska 2001), confocal microscopy (Reddy et al. 2004) and, most recently, multi-angle imaging via confocal microscopy (Fernandez et al. 2010). To date, for almost a century, the method of
indirect cell lineage tracing known as clonal analysis has provided the most reliable and overall observations about cell lineage activity in plants (Korn 2001). On the basis of our protocol for
reconstruction, we have developed a further protocol for total cell lineage tracing which is described below. We believe that our approaches are capable of filling the methodical and technical gaps
highlighted above.

In this work, we applied both our protocols to recreate the complex shape of the plastids and mitochondrial network in the tuber parenchyma cells of \textit{Corydalis bracteata} (Steph.) Pers.
(Fumariaceae) and to reconstruct the cellular architecture and developmental history of the embryo and shoot apex of \textit{Calla palustris} L. (Araceae).

\section*{MATERIALS AND METHODS}

\subsection*{Plant material}

\textit{Corydalis bracteata} plants were collected in the Botanical Garden of the Russian Academy of Sciences in St-Petersburg, Russia, in April, 2008. Mature 3-4 year old tubers (about 1.5 cm in
diameter) were chosen for sampling. \textit{Calla palustris} seeds and seedlings were collected in natural habitats in Trubnikov Bor, Leningrad region, Russia, during 2006 to 2008.

\subsection*{Transmission electron microscopy}

For organelle reconstructions, cells of tuber phloem parenchyma of \textit{C. bracteata} were analyzed. Immediately after collection, tuber pieces were fixed for 48 h with a mixture of 2.5 \%
paraformaldehyde and 2 \% glutaraldehyde in 0.1 M phosphate buffer, pH 7.4, and then washed three times with fixative buffer followed by 12 h of fixation at 4°C with 2 \% osmium tetroxide in 0.1 M
phosphate buffer, pH 8. Tuber pieces were then washed three times with fixative buffer, dehydrated using a 30 \%, 50 \% and 70 \% ethanol series and then stained with 2 \% uranyl acetate for 2 h at
room temperature. They were subsequently dehydrated using 70 \%, 95 \% and 100 \% ethanol and 100 \% acetone, and infiltrated in Epon-Araldit M resin (Fluka, Switzerland) using a 1:5, 1:4, 1:3, 1:2
and 1:1 (resin:acetone) series with 1 h of incubation in each solution. Samples were transferred to pure resin, cast into 0.4 ml capsules, and polymerized at 60°C for 72 h. 
Semi-thin and thin sections were cut on an Ultracut E (Reichert, Germany) ultratome using a diamond knife (Diatome, Switzerland). Thin sections were stained with 4 \% uranyl acetate in water and
Reynold's lead citrate and examined at 60 eV and documented in a Hitachi-H600 (Hitachi, Japan) transmission electron microscope.

\subsection*{Light microscopy}

Immediately after collection, plant materials of \textit{C. palustris} were fixed in FAA (formalin 40 \%, alcohol 70 \% and acetic acid 98 \%, 7:100:7). After dehydration in the alcohol series,
acetone and chloroform, the samples were embedded in paraffin and sectioned on a Microm 325 (Carl Zeiss, Germany) microtome. After deparaffinization in xylene, the sections were rehydrated and stained
in Foelgen stain, alcyan blue and Erlich hemathoxylin. Then sections were mounted in Mowiol 4-88 (Fluka, \#81381, Switzerland) and observed in an Axioplan 2ie (Carl Zeiss, Germany) microscope. Optical
sections were made using a 63x/1.4 Plan-Neofluar (Carl Zeiss, Germany) objective. Sections were imaged using a Nikon D70 (Nikon, Japan) digital camera and by means of Nikon capture 1.4.2. (Nikon,
http:$\sslash$www.nikon.com) software saved as .tiff files.

\subsection*{Organelle and tissue reconstruction}

Digital images were processed and structures were contoured using the Adobe Photoshop CS (Adobe, http:$\sslash$www.adobe.com) graphic editor and Wacom PTZ-1230d (Wacom Company, Japan)
drawing tablet. Location along the $Z$ axis, alignment, surface generation, material assignment visualization and cell lineage tracing were carried out using 3DSMax 7.0 (Autodesk,
http:$\sslash$usa.autodesk.com) and Blender 2.49 (Blender Foundation, http:$\sslash$www.blender.org) 3D graphic editors. 

\section*{RESULTS}

\subsection*{2D image acquisition}

The tuber phloem parenchyma cells of \textit{C. bracteata} were cut into 50 serial sections, 70 nm thick for TEM. The stack of section images was stored as .tiff format files with a resolution in the
$XY$ plane of about 0.33 x 0.33 nm per pixel (Fig. 1a). The \textit{C. palustris} embryo was cut (for LTM) into 9 serial sections inclined to the vertical axis of the embryo, 5 µm thick, and then each
of them was scanned via 800 nm along the $Z$ axis. 58 optical sections were obtained from these nine sections (Fig. 1b). The \textit{C. palustris} shoot apex was cut (for LTM) into two serial
transversal sections, 12 µm thick, and then each of them was scanned with a step of 1 µm along the $Z$ axis. 29 optical sections were obtained from these two serial sections (Fig. 1c). In the last two
cases, this was a little more than should be expected (24-26). According to N. White and colleagues (White et al. 1995), these aberrations should be considered as the axial focus errors caused by the
specimen refraction index being higher than that of the embedding medium and a $Z$-correction factor should be applied for each physical section of known thickness. Obviously, the greater thickness of
sections for LTM significantly influences the appearance of such aberrations. The stacks of images of optical sections were stored as .tiff files with an $XY$ resolution of about 50 x 50 nm per pixel.
An extremely high resolution in the $XY$ plane was used for more accurate recognition of cell walls and membranes of organelles.

\subsection*{Contouring}

The first step in the image processing is a contouring of object boundaries which are then saved as curves or vector linear objects, known as splines. After these operations, the object of interest
becomes a set of contours or splines. Each contour is a curve, often closed and described by a set of points associated with tangent vectors. The object boundaries can easily be recognized
automatically if the background-structure transition coincides with the threshold of brightness, hue or other pixel properties. For instance, images obtained by fluorescence and confocal approaches
are actually bicolor and conform well to the automation of contouring also known as segmentation (Fiala and Harris 2002; Fiala 2005). Hence, there is a big range of commercial and open-source software
available for the segmentation of such images. Nevertheless, ever since classical anatomical drawing, tracing an object boundary manually still gives the most reliable results in complex situations
even though it is time-consuming. Manual segmentation is always a “golden standard” for structure recognition for “virtual” microscopy as well (Khairy and Keller 2011). To reduce the time required for
manual tracing, a drawing tablet was used.

Boundaries of organelles of \textit{C. bracteata} in the images obtained from TEM were recognized by dyeing the membrane black (Fig. 1d). All the section images of the embryo and shoot apex of
\textit{C. palustris} were multicolored, so the boundaries between cells were recognized either by cell wall tracing or by cytoplasm discontinuity, or by counting cell nuclei (Fig. 1e, 1f). Using a
raster 2D computer graphic editor, Adobe Photoshop CS (Adobe, http:$\sslash$www.adobe.com), the boundaries of organelles and cells were outlined with the curves, which were then exported in a vector
format .ai. The same result was achieved using the GIMP graphic editor (http:$\sslash$www.gimp.org) except that the format of the curve export was .svg (see Table 1). Among microscopy-oriented
software, the Reconstruct package should be mentioned as an environment which allows not only the contours to be traced both manually and automatically but also complex images of large sections to be
constructed from fragments and aligned along the $Z$ axis (Fiala and Harris 2002; Fiala 2005).

\end{multicols}

\begin{figure}[htb!]
\centerline{\includegraphics{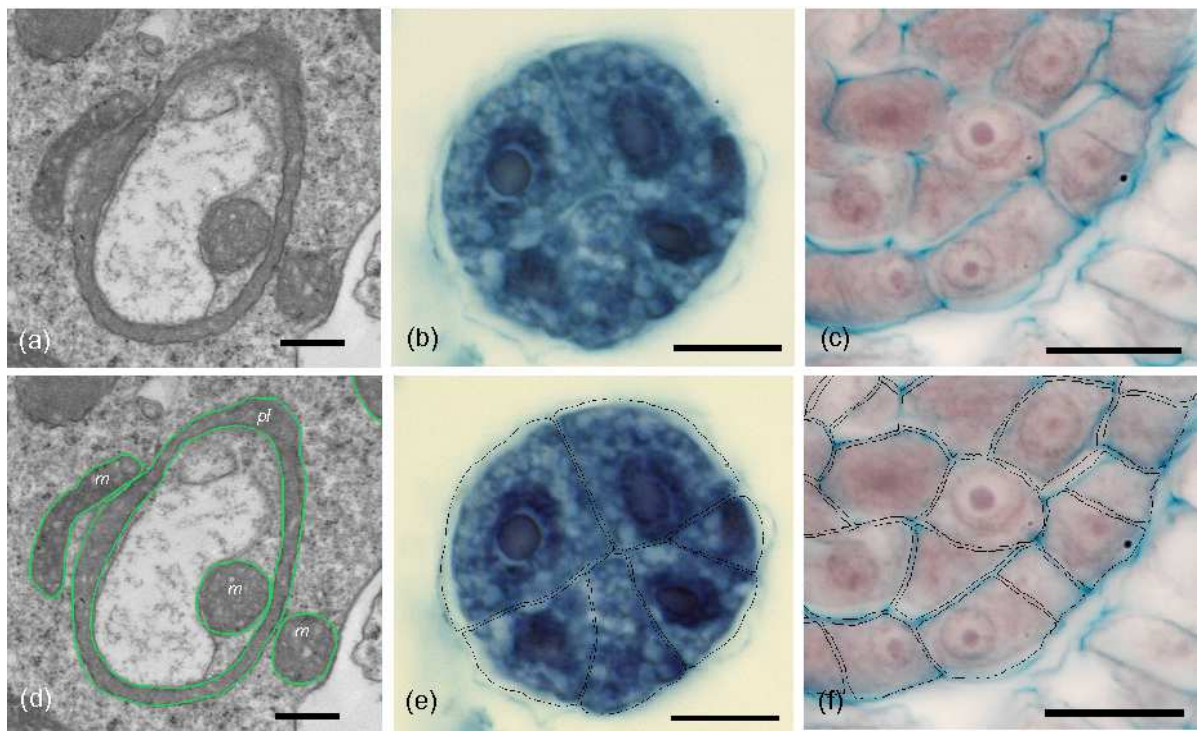}}
\caption{\footnotesize Images of serial sections (a-c) and their contouring (d-f).\\
\textbf{(a)}, \textbf{(d)} – plastids (\textit{pl}) and mitochondria (\textit{m}) of \textit{Corydalis bracteata} tuber parenchyma cell. Bar = 500 nm.\\
\textbf{(b)}, \textbf{(e)} – \textit{Calla palustris} embryo. Bar = 10 µm.\\
\textbf{(c)}, \textbf{(f)} – \textit{C. palustris} shoot apex. Bar = 10 µm.}
\label{Fig1}
\end{figure}

\begin{multicols}{2}

\subsection*{Location along the $Z$ axis}

The location of contours or splines along the $Z$ axis and their alignment in the $XY$ plane recreate the shape of the objects of interest in 3D space. The 3D vector graphic editors 3DSMax (Autodesk,
http:$\sslash$usa.autodesk.com) and Blender (Blender Foundation, http:$\sslash$www.blender.org) were applied in our work with the same success for the correct location of the contours in the virtual
3D space. The contours of organelles of \textit{C. bracteata} were consequently imported and moved along the $Z$ axis according to the thickness (70 nm) of their physical sections (Fig. 2a). The
contours of cells of the embryo and shoot apex of \textit{C. palustris} were successively located along the Z axis with respect to the step between the optical sections, the thickness of physical
sections and the $Z$-correction factor (Fig. 2b).

The automation of the representation of contours in 3D is usually based on the reading of $Z$-data enclosed in the special image formats for optical (confocal) sections. The majority of special
software for serial microscopy works with a uniform distance between physical sections for the $Z$ axis. Only some of them, like Reconstruct (Fiala 2005) and IMOD (Kremer et al. 1996), can perceive
different distances between each section.

\subsection*{Alignment}

The alignment of object contours in the $XY$ plane involves applying translational and rotational transformations to them in order to align sections by fiducial points and maximum similarity.
Misalignment, caused by section deformation like stretching or folding in the plane or section curving, can be corrected by scaling along the deformation direction; however, the absence of such
defects indicates the quality of the source material. In the case of lower optical resolution and greater section thickness, typical of low optical magnification, the manual method of alignment is the
most suitable (Hofstadler-Deiques et al. 2005).

Contours of organelles made from serial sections of \textit{C. bracteata} cells were aligned manually by their maximum similarity (Fig. 2a). Contours in the stack of optical sections of the \textit{C.
palustris} embryo and shoot apex did not need alignment in the $XY$ plane, because the format of the export/import files (.ai) reproduced their position completely in the source images (Fig. 2b).
Physical (serial) sections of \textit{C. palustris} samples were aligned manually by top and bottom optical sections of consequent physical sections by maximum similarity and fiducial points of
splines. Then the alignment was verified by fiducial surfaces of cells. Quick alignment by surface was achieved when the surface was easily traced by contours of optical sections without surface
generation (Fig. 2c). After the alignment of all the optical sections had been accomplished, most mistakes in the object boundary tracing were indicated and corrected.

Section alignment in TEM-oriented software is conventionally automated by means of the approach of maximum similarity between section images and is usually performed before object contouring (Fiala
2005; Kremer et al. 1996). To date, the alignment by fiducial surfaces of the object applied in our work can only be carried out manually in conventional 3D vector graphic editors.

\subsection*{Surface generation}

The last stage of the reconstruction of the object of interest is the generation of a mesh and surface over the aligned set of the object’s contours. The splines of the object, which are parallel in
the $XY$ plane, are point-to-point connected to each other with a set of vertical or longitudinal splines, such that a mesh of triangles or tetragons is generated. In order to avoid shape twisting, it
was necessary to provide the locally equal density of points along the successive contour splines. Proper (without twisting) triangular and tetragonal faces of the mesh of splines serve as a skeleton
for surface generation in arbitrary 3D graphic software. After generation of its surface, the object of interest acquires its own visible shape which enables it to be analyzed, measured and
manipulated.

During the manual and automatic mesh construction of \textit{C. palustris} shoot apex cells and \textit{C. bracteata} plastids, three types of longitudinal connection of contours parallel in the $XY$
plane were found with respect to the object`s surface properties. In the first case, the longitudinal splines smoothly connected points of the successive section splines (Fig. 2d). Secondly, in the
case of surface bifurcation, the longitudinal splines bifurcated to connect the surface lobes and to cover the saddle zone (Fig. 2e). Thirdly, when a contour was the last or first in a stack, the
longitudinal splines were fused to close the shape opening (Fig. 2f). Proper selection of the connection type was highly critical to the identification of the object. In particular, if the connection
was made automatically, it was important to check for erroneous structure merging, overlapping or fragmentation.

The conventional 3D vector graphic editors appeared to be the most convenient environment to carry out these operations both manually and automatically (Table 1). In the present work, 3DSMax
(Autodesk, http:$\sslash$usa.autodesk.com) and Blender (Blender Foundation, http:$\sslash$www.blender.org) were applied equally. Surface generation was carried out on the skeleton of splines by means
of the “Surface” modifier and on mesh edges by means of faceting with the “Subsurface” modifier in the respective programs. 

The most frequently used approach of automatic surface generation in microscopy-oriented software is the “marching cubes” method (Lorensen and Cline 1987; Newman and Yi 2006), based on the
verification of surface points in each volume element (voxel) of the object. Perhaps this direct “voxelization” could explain the typical absence of vector formats for data import/export in
microscopy-oriented programs. To our knowledge, only a few types of software, namely Reconstruct (Fiala 2005), IMOD (Kolibal and Howard 2008), Imaris (Bitplane, http:$\sslash$www.bitplane.com) and
Amira (Visage Imaging, http:$\sslash$www.amira.com), provide this property (Table 1).

\end{multicols}

\begin{center}
\centerline{\includegraphics{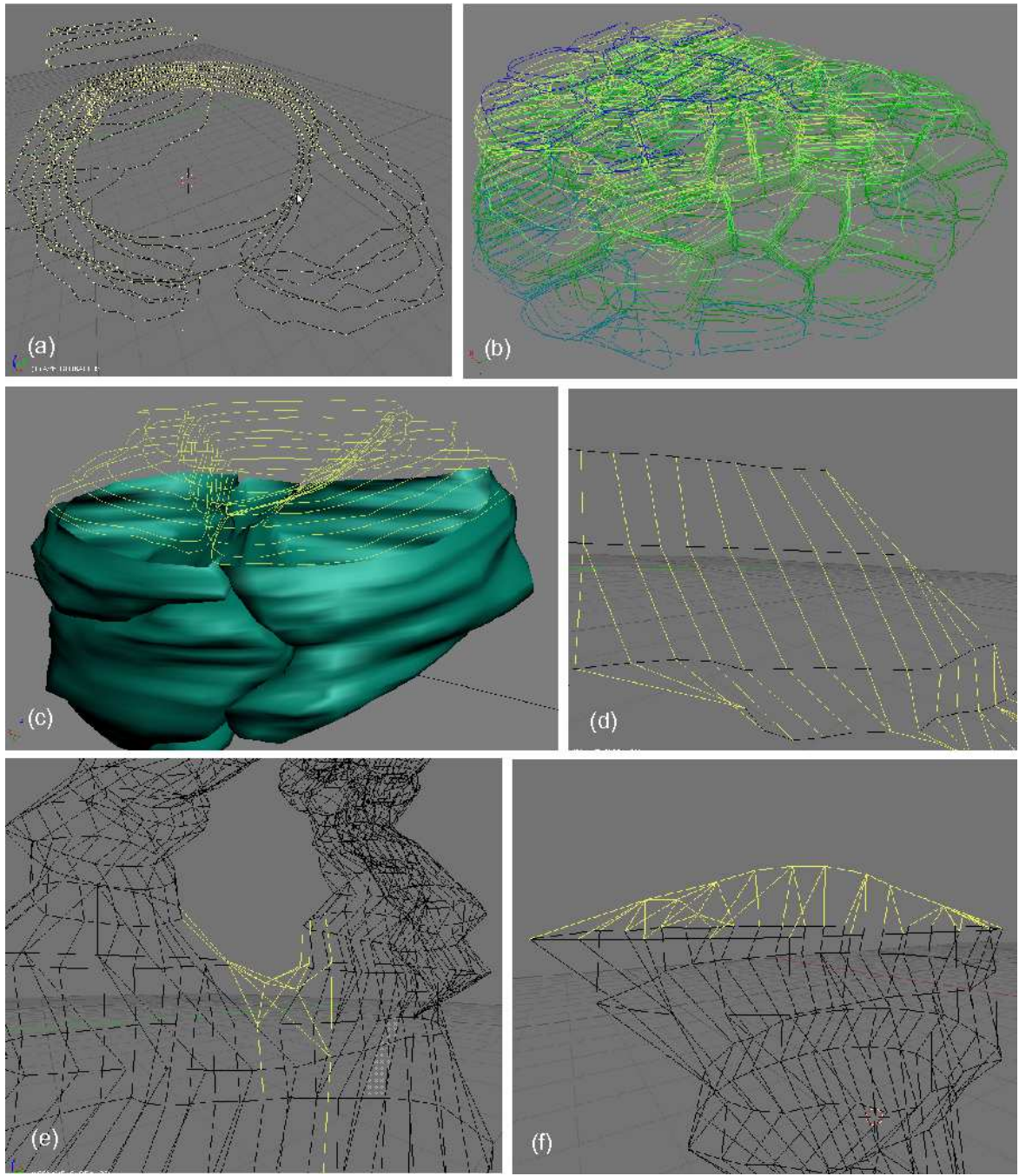}}
\captionof{figure}{\footnotesize Location and alignment of contours and mesh construction in 3D.\\
\textbf{(a)} – contours of serial sections located along the $Z$ axis according to their thickness and aligned by maximum similarity. Plastids of \textit{C. bracteata}.\\
\textbf{(b)} – pre-aligned stack of contours of optical sections located along the $Z$ axis according to the step of vertical scanning. \textit{C.~palustris} shoot apex.\\
\textbf{(c)} – alignment of stack of contours by (fiducial) surfaces of already reconstructed cells. \textit{C.~palustris} embryo.\\
\textbf{(d-f)} – mesh construction upon contours. Longitudinal splines shown in yellow. Mitochondria of \textit{C. bracteata}.\\
\textbf{(d)} – planar connection of contours.\\
\textbf{(f)} – construction of saddle zone in the case of surface (mitochondria) bifurcation.\\
\textbf{(c)} – closing of shape opening.}
\label{Fig2}
\end{center}

\begin{multicols}{2}

\subsection*{Surface editing}

The real surface of the object might have sharp edges, like the cell wall of the plant epidermal or parenchymatous cells for instance, so the point-to-point connection of curves corresponding to
object boundaries must reflect this. Moreover, especially in the case of erroneous surface generation or contouring, the generated surface needs editing. This involves moving the surface mesh points
into the proper position in 3D and changing the sharpness of the surface around definite points. Any conventional 3D computer graphic software can do this, while only some of the microscopy-oriented
programs have this option. Some microscopy-oriented software, like Reconstruct and IMOD (Kremer et al. 1996; Fiala 2005), are suitable for manual spline or surface editing also (Table 1). However, the
remaining special programs for microscopy are unfortunately not. Automation of the surface (“isosurface”) generation in this software is based on the threshold visualization principle, similar to that
of automatic contouring with the same limitations. Thus, a surface automatically generated by means of most software is usually uniform and either sharp or smooth. The latter is preferable in order to
achieve a better visual quality of the surface. As a result, less structural information can be obtained from such models. 

For the proper representation of surface edges in \textit{C. palustris} cells, the edges were traced through successive section splines. During this process, two types of surface edge appearance were
encountered. (1) If the surface edge lies in the $XY$ plane between definite points of the section spline (Fig. 3a, red), the longitudinal (vertical) spline should have the points of sharp curve
bending when crossing this segment. (2) If the surface edge crosses the $XY$ plane (Fig. 3b), it is already reflected in the sharpness of the section contours, i.e. there are points of sharp spline
bending. In this case, the longitudinal spline should connect such points of consequent contours with respect to the corresponding surface edges. In both cases, the surface edging was due to the
influence of neighboring structures. For the more detailed significance of the cell surface edges, see below.

\end{multicols}

\begin{figure}[htb!]
\centerline{\includegraphics{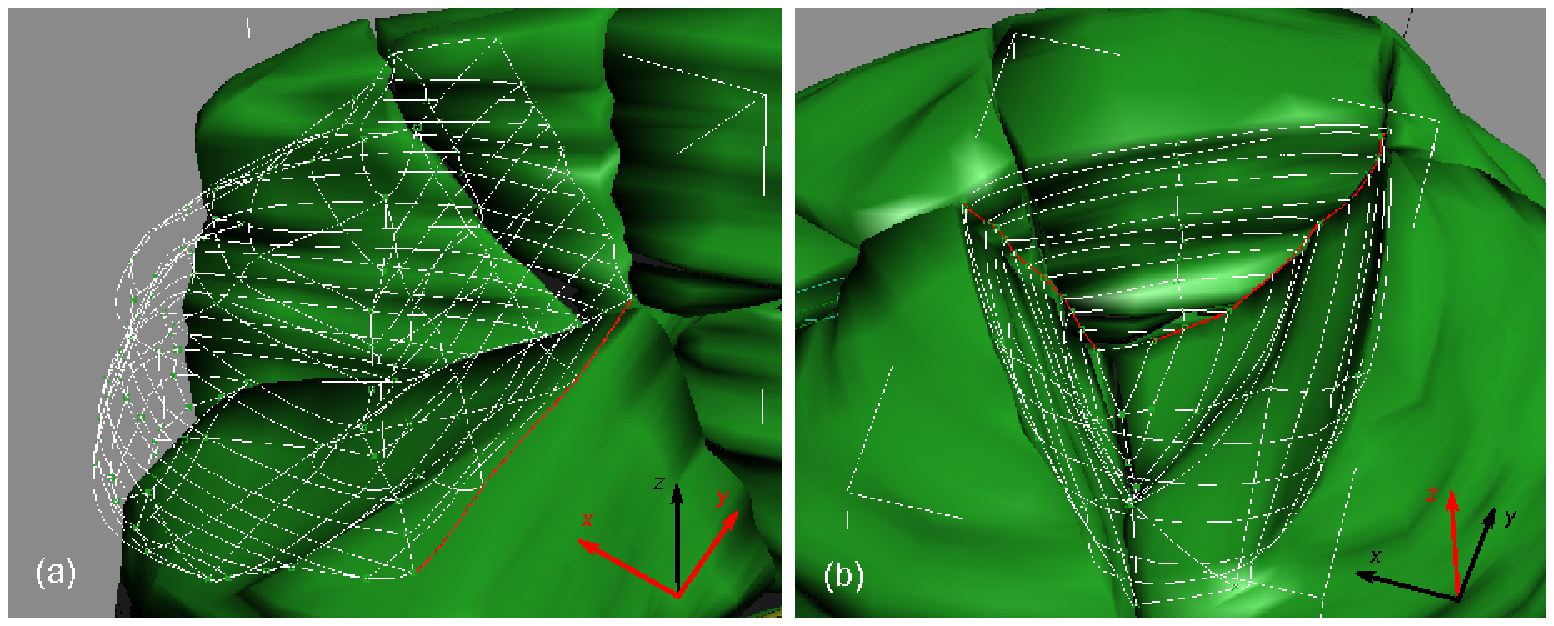}}
\caption{\footnotesize Tracing of cell surface edges with respect to neighboring structures. Surface edges correspond to the curves (shown in red) which connect in the mesh the points of sharp curve
(spline) bending. Cells of \textit{C. palustris} embryo.\\
\textbf{(a)} – surface edge is a part of the contour and lies in the $XY$ plane.\\
\textbf{(b)} – surface edges cross the $XY$ plane and contours.}
\label{Fig3}
\end{figure}

\begin{multicols}{2}

\subsection*{Material assignment and visualization}

The creation and assignment of virtual materials to the reconstructed objects and their visualization are necessary operations for the adequate transfer and perception of volume information about the
object by the observer. Materials can be assigned only to objects with a surface. Each material is a set of properties, which describe diffuse color, opacity, maps of reflection and refraction etc. As
a rule, the conventional 3D graphic editors and renderers are able to recreate almost any material that exists in nature. Reconstructed objects might be numerous and very complex; they might occur one
inside another and, like the organelles and other subcellular structures, might not have their own natural color due to the diffraction limitation. In the latter case, artificial material must be
created for the proper visualization of the structures investigated.

Using Blender (Blender Foundation, http:$\sslash$www.blend er.org) and 3DSMax (Autodesk, http:$\sslash$usa.autodesk.com) with Mental Ray (Autodesk, http:$\sslash$usa.autodesk.com) renderer software,
some materials of similar type but with different diffuse colors and relief maps were assigned to the plastids and mitochondria of \textit{C. bracteata} and to cells of \textit{C. palustris}. Each
material was composed of the same map of reflection and shaded as being metal-like (Fig. 4a-c). Some degree of self-luminescence was assigned to the material of stem cells in the shoot apex of
\textit{C. palustris} (Fig. 4c). In order to achieve the photo-realistic and voluminous visualization of the object with the material assigned, special illumination was required in the scene of the 3D
graphic renderer. From one to three lighting sources were placed in the scene with the objects to achieve a better emphasis of the structures of interest (Fig. 4d).

Microscopy-oriented software usually possesses some types of default materials with a limited set of properties such as diffuse color, opacity and self-luminescence (Table 1). Unfortunately, the
volume models reconstructed in this way only look good when animated on a black background. Nevertheless, this visualization is rather fast and well suited to real-time object observation. A good
solution to the problems mentioned is to export the model to any final renderer which is supported in some microscopy-oriented programs (Table 1). Noticeably, such software usually supports only
default illumination with the exception of BioImageXD (BioImageXD group, http:$\sslash$www.bioimagexd.net) which possesses a range of rendering and lighting options.

\end{multicols}

\begin{figure}[htb!]
\centerline{\includegraphics{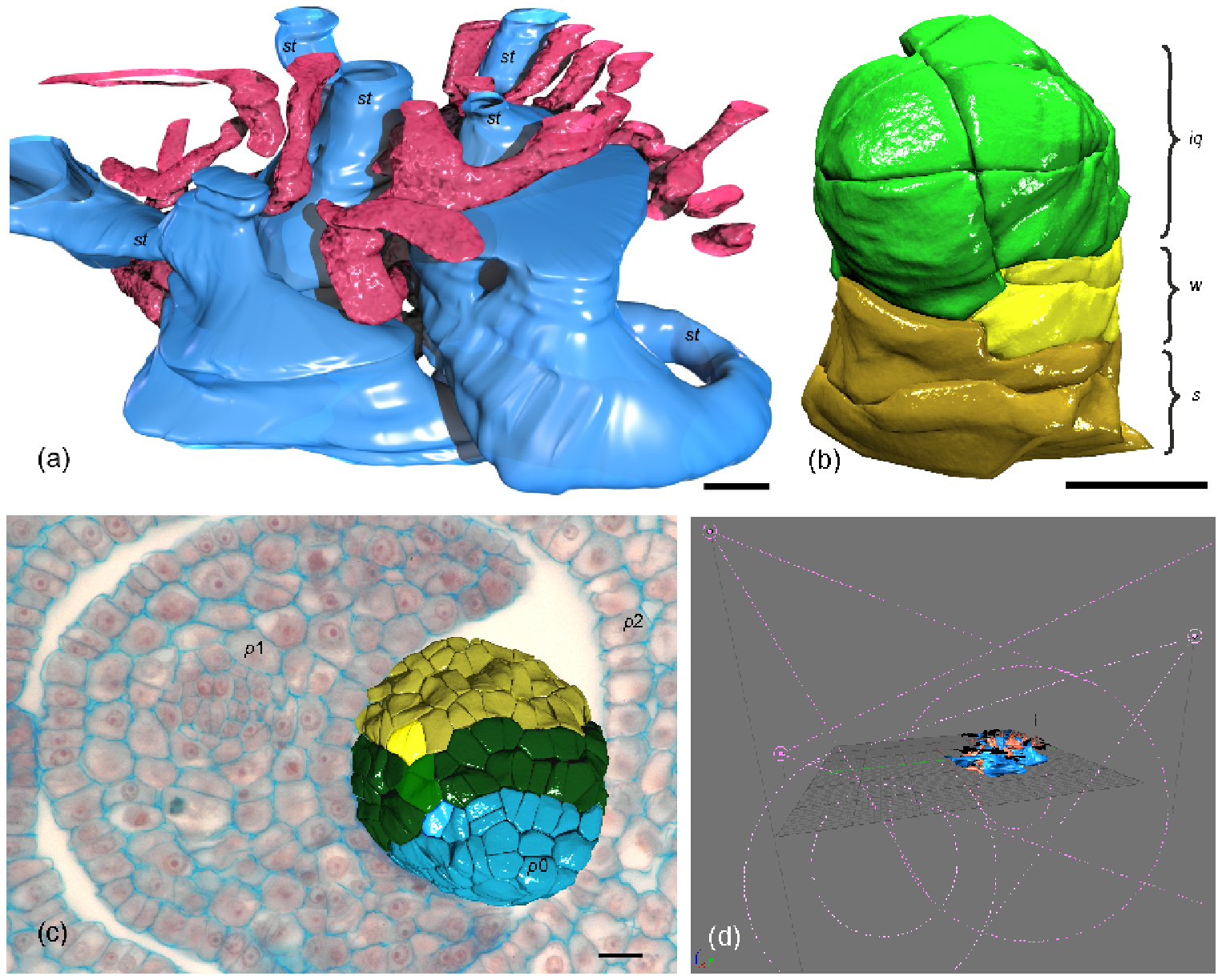}}
\caption{\footnotesize Visualization of the 3D structure.\\
\textbf{(a)} – plastids (blue) and mitochondria (red) networks packed inside the tuber parenchyma cell of \textit{C. bracteata}. \textit{st} – stromule. Bar = 500 nm.\\
\textbf{(b)} – \textit{C. palustris} embryo. Three basic cell lineages shown with different colors. \textit{iq} – quadrants' initial (green); \textit{w} – wedge-shaped cell (yellow); \textit{s} –
suspensor (brown). Bar = 10 µm.\\
\textbf{(c)} – \textit{C. palustris} shoot apex. $p0$-$2$ – leaf primordia numbers. Bar = 10 µm.\\
\textbf{(d)} – illumination of 3D scene. Lighting sources shown as red spots with cone of illumination.}
\label{Fig4}
\end{figure}

%
%

\begin{table}[h!]
\caption{\footnotesize Software features with respect to 3D reconstruction based upon serial sections. A – automatic, M – manual. The central horizontal bold line separates the conventional computer
graphic software from microscopy-oriented software.}
\centerline{\includegraphics{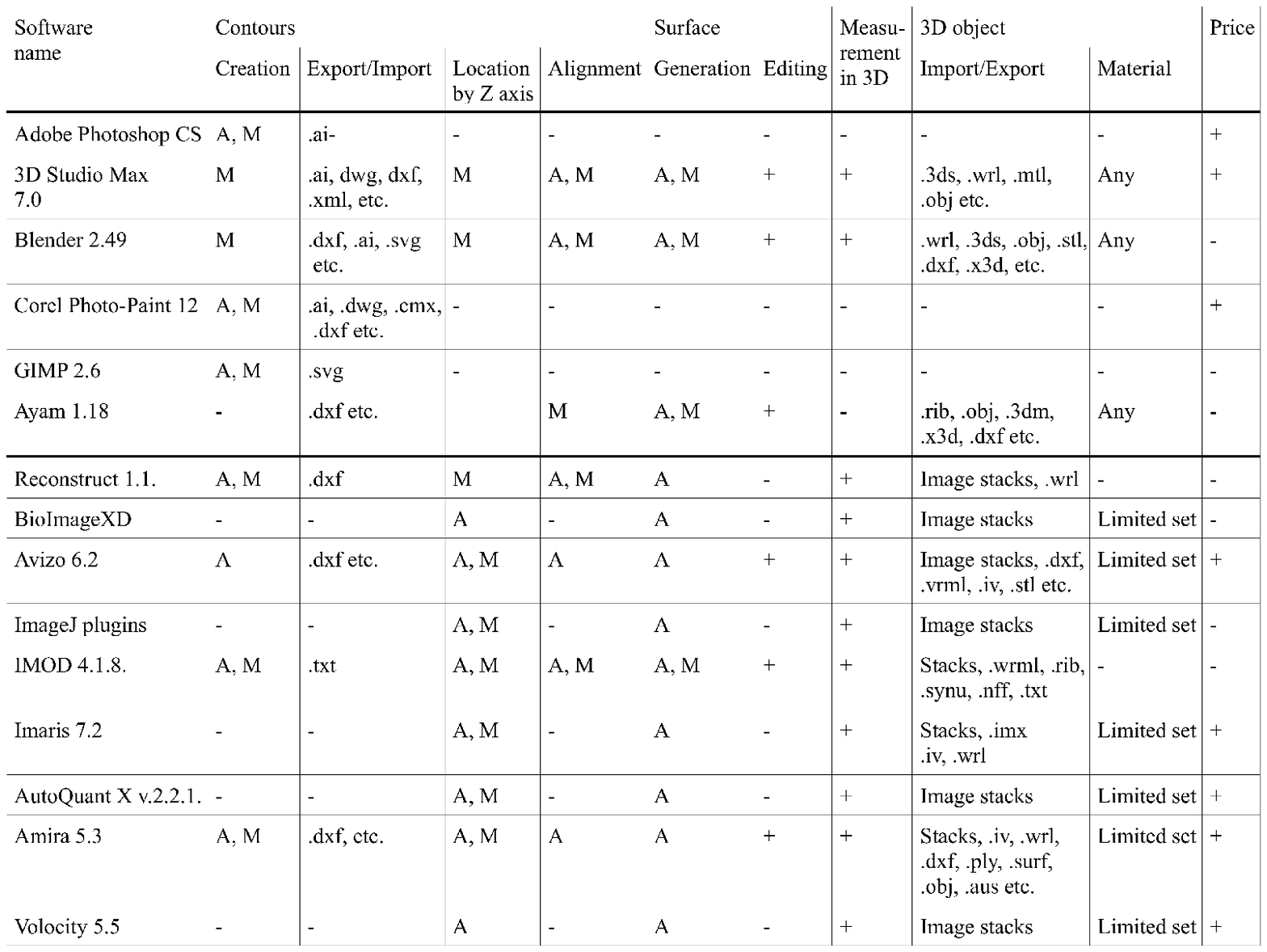}}
\label{Tab1}
\end{table}

\begin{multicols}{2}

\subsection*{Cell lineage tracing}
 
This procedure was carried out only with samples of \textit{C. palustris}. The shape of the cell wall and the adjacency of cells are the main information sources for the
reconstruction of the developmental history of tissue and cell lineage tracing. The algorithm of cell genealogical tree reconstruction developed in this work involves three steps: (I) the search for
any last division or any two sister cells, (II) the successive revelation of their sister merophytes from previous generations and (III) tracing all of the indirectly related cells inside the sister
merophytes found.

Genealogical relationships between cells have been estimated by means of cell shape analysis based on R. Korn’s approach (1974) developed for tissues with directed growth. In the present work, owing
to the 3D representation of cellular architecture, it became possible to analyze tissues with non-directed (multi-directed) growth. 

Let us consider the basic theoretical fundamentals and cell shape properties that are sufficient and reliable for cell lineage tracing. The question to be resolved here is as follows. The tissue
fragment consists of $n$ cells specifically spatially arranged. One should trace back all cell divisions occurring in this tissue sample in the course of the appearance of these exact $n$ cells. These
$n$ cells were generated after at least $n - 1$ cell divisions, beginning with the single common ancestral cell independent of the genealogical tree complexity, since all plant cells remain connected
after division (Fig. 5a, 5b). If we can claim that no cells of the tissue sample were missed or resorbed completely, the genealogical tree of this tissue fragment consists of  $2n-1$  vertices,
whose n vertices are pendant and correspond to the “visible” cells of the fragment, and  $n-1$  cells are vertices with degree tree, i.e. branching points of the tree, with the exception of the root
vertex (degree two) that corresponds to the common mother cell. It is easy to see that each  $n-1$  cell division that occurred in the past is mapped onto the  $n-1$  cells of the tissue fragment.
This mapping is denoted by the direction of the tree edges (Fig. 5a). If some cells are missed or completely resorbed , the cell genealogy of the tissue fragment can be only partially reconstructed. A
set of the relatively independent cell lineages with their genealogy will be revealed instead of the single genealogical tree. These partial genealogies can be coarsely united via the common root
vertex since the cells of relatively independent lineages are connected (Fig. 5b). Hence, this mapping carries the complete information about preceding cell divisions of any single cell from the set
of $n$ cells of the tissue fragment. This information is specifically inscribed into the spatial arrangement and shape of cells of plant tissue. More precisely, it is a genealogical distance between
any pair of $n$ cells and their common ancestral cell, measured by the number of cell divisions. Thus, any two sister cells are equally distanced from their mother cell by one cell division each,
while any two non-sister cells (without respect to missed or resorbed cells) are arbitrarily distanced from the common mother cell but for no more than  $n-1$  cell generations.

\end{multicols}

\begin{figure}[!htb]
\centerline{\includegraphics{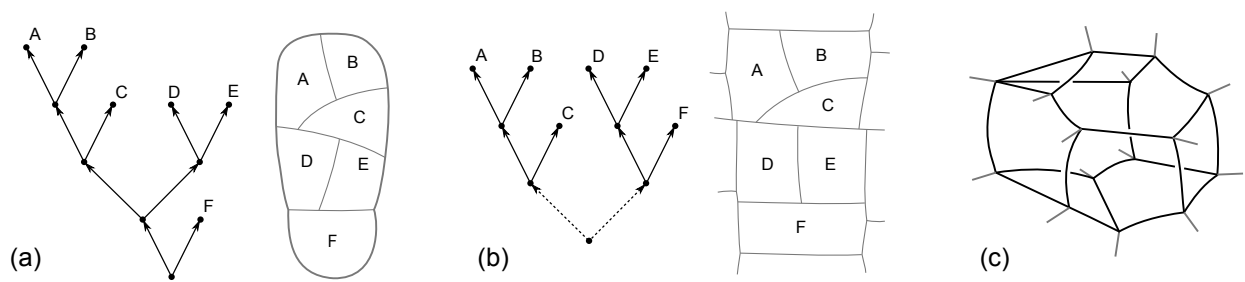}}
\caption{\footnotesize Fundamentals of cell lineage tracing based on geometry of cells.\\
\textbf{(a-b)} – genealogical trees and tissue fragments. Relation between the number of cells $n=6$ in a tissue fragment and the number of branching vertices of their genealogical trees $n-1=5$.\\
\textbf{(a)} – tissue fragment without cells missed (embryo).\\
\textbf{(b)} – tissue fragment with inevitable missing of cells outside of three celled merophytes. Dotted edges correspond to the partial recreation of cell genealogy in the sample. Indeed, the
mother cell of this tissue fragment is more distanced from merophytes “$ABC$” and “$DEF$”.\\
\textbf{(c)} – natural polyhedron-like structure of cell walls in plant tissue. The edges (black) of a cell are the boundaries of facets shared with neighboring cells whose fragments of edges are
shown in gray.}
\label{Fig5}
\end{figure}

\begin{multicols}{2}

\textbf{I. Search for sister cells.} Consider several basic properties of cell shape and cell spatial arrangement; these carry information about cell genealogy in the order of their importance for
genealogy tracing. (1) Only two adjacent plant cells are capable of being sister cells in the tissue. Two non-adjacent cells in a tissue are genealogically distanced from being sister cells for at
least  $k+1$ cell generations, where $k$ is a minimal number of cells between these two non-adjacent cells. For instance, look at cells “$A$” and “$B$” in Fig. 5a. This property is easily
observable, however k can be numerically estimated as a spatial distance between cell vertices in a graph of spatial adjacency introduced upon the cellular spatial arrangement (Rudskiy et al. 2011).

(2) Consider the geometric properties of a cell in plant tissue. All cells in a tissue are adjacent to each other along the contiguous cell wall or the facet. Every facet of a cell is adjacent to
other facets of the same cell or to the facets of neighboring cells along its boundaries, i.e. edges (see Fig. 5c). In plant tissues, each cell is adjacent to 10 – 15 cells on average (Korn 1974). 

Two adjacent cells are more likely to be sister cells as the edges of their common facet do not influence the shape of the union of these two cells. In other words, any two adjacent cells are
considered as sister cells with a maximal probability if their united shape looks quite like one cell, i.e. locally convex and smooth without sharp concavities or prominences along the edges of the
common facet. Local convexity and smoothness of the surface of cell union can be easily measured visually with the help of 3D graphic editors or estimated in the form of the surface curvature
coefficients in the neighborhood of the boundary of the united cells (Fig. 6a, 6b), although that needs additional computational work and regarding cell surfaces as closed two-dimensional manifolds.

(3) Two adjacent cells with a locally convex and smooth external boundary of their union are more likely to be sister cells as their common facet has the least surface curvature (Fig. 6a, c). This
property is natural since it corresponds to the well known interpretation of L. Erera's law of cell divisions where the cell wall between dividing cells is the minimal surface (reviewed in Priestley
1929; Sou\`eges 1936). In the course of cell divisions and growth, even a slight difference in growth patterns, inevitable in multi-directly growing plant tissues, leads to the distortion of the facet
and bending of the edges between sister cells, thus providing a natural time measurement for the newly generated cell wall.

\end{multicols}

\begin{figure}[htb!]
\centerline{\includegraphics{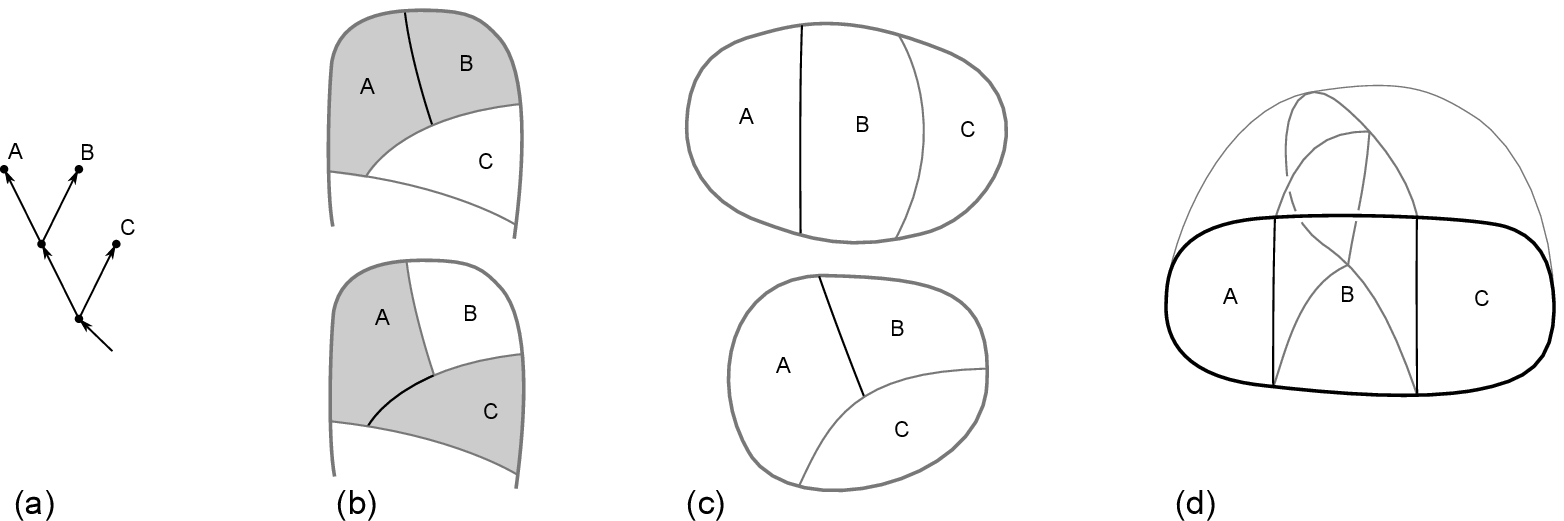}}
\caption{\footnotesize Search for sister cells in tissue fragment.\\
\textbf{(a)} – sample genealogical tree for the following cases.\\
\textbf{(b)} – two adjacent cells are more likely to be sister cells if the edges of their common facet do not influence the shape of the union of these cells. Comparison of two (from three) possible
unions of pairs of cells “$A$”, “$B$” and “$A$”, “$C$” (shadowed) with respect to their common facets (black).\\
\textbf{(c)} – two adjacent cells are more likely to be sister cells if their common facet has the least surface curvature (black). \\
\textbf{(d)} – possibly irresolvable situation in 2D (black section) almost always has a unique and reliable solution in 3D.}
\label{Fig6}
\end{figure}

\begin{multicols}{2}

\textbf{II. Search for sister merophytes.} In general, two sister cells were determined unequivocally in all tissue samples (Fig. 7a-c). The cell or its descendants (merophytes) most closely related
to the first two sister cells determined were considered the result of some previous division. Such sister cells (merophytes) were called sister cells (merophytes) of previous generations according to
the order of their relationship. The search for a sister merophyte was implemented analogously to the principles (1) – (3) listed above for sister cells (Fig. 7d, 7e).

Successive iteration of this search enabled the sister merophytes (cells) of higher and higher orders to be found. In other words, a chain of points of bifurcation of the genealogical tree was
established (Fig. 7c, 7e). 

A situation could arise where more than two adjacent merophytes (cells) have a smooth common shape and an equal curvature of their common facets. This could be resolved by verifying all the sister
merophytes of each candidate. In a 3D representation of a tissue sample, this is easy to implement (Fig. 6a, d). If more than two merophytes still look like sister ones, it is possible that all these
merophytes are closely related, but perhaps the two most related merophytes should have a similar volume which corresponds to J. Sachs' and O. Hertwig's law of cell divisions, concerning the
equivalence of masses or volumes of dividing cells (reviewed in Priestley 1929; Sou\`eges 1936). There are also several other natural markers of cell division sequence like, for example, cell wall
thickness arising from the coordinated laying down of the cell wall on the boundary of the merophyte (L\"uck et al. 1994).

\end{multicols}

\begin{center}
\centerline{\includegraphics{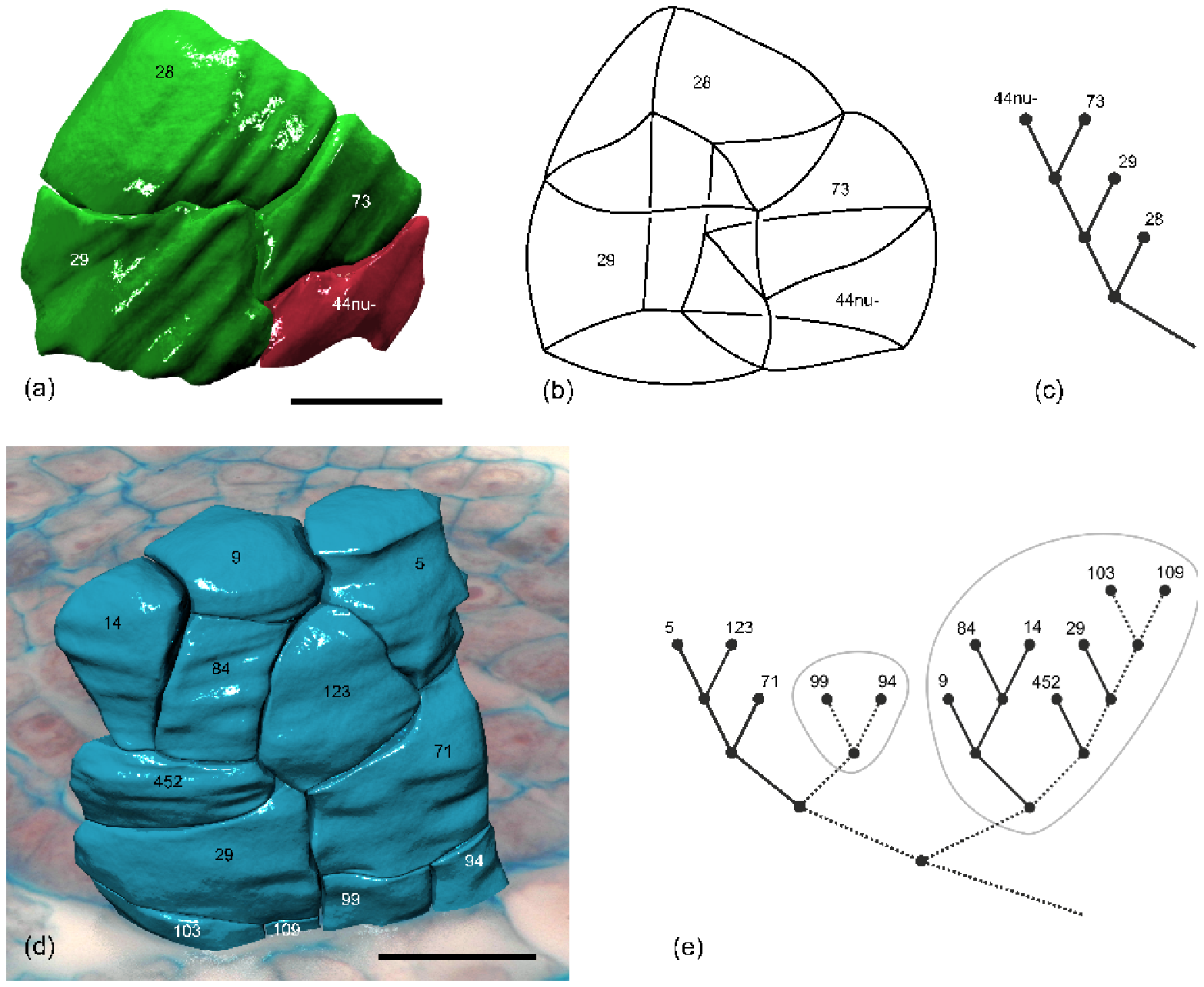}}
\captionof{figure}{\footnotesize Cell lineage tracing example. Each cell is indicated by an individual number. Bar = 10 µm.\\
\textbf{(a-c)} – reconstruction of the main sequence of cell divisions.\\
\textbf{(a)} – 3D visualization of the fragment of \textit{C. palustris} embryo.\\
\textbf{(b)} – scheme of cell edges and facets.\\
\textbf{(c)} – genealogical tree for two sister cells “73” and “44$nu$-”. Cells “29” and “28” are sister cells of previous generations.\\
\textbf{(d-e)} – reconstruction of the main sequence of cell divisions for cells “5” and “123” and full genealogy of the tissue fragment. Some sister merophytes are highlighted.\\
\textbf{(d)} – 3D visualization of the fragment of \textit{C. palustris} shoot apex.\\
\textbf{(e)} – genealogical tree for each merophyte.}
\label{Fig7}
\end{center}

\begin{multicols}{2}

\textbf{III. Total genealogy recreation.} The reconstruction of the whole genealogical tree involves the iteration of the first two steps I and II (i.e. finding the two sister cells etc.) inside each
merophyte related to the two primary determined sister cells (Fig. 7d, 7e). Complete cell lineage tracing is possible only when the whole organism is investigated (the embryo, for instance). In the
case of tissue samples, only the relatively independent cell lineages can be found since a part of the merophytes remained incompletely reconstructed due to the tissue sample boundaries. The
genealogical relations of these merophytes are indicated by dashed lines (Fig. 7e).

The operations required for the tracing of cell lineages were carried out manually using conventional 3D editors like 3DSMax (Autodesk, http:$\sslash$usa.autodesk.com) and Blender (Blender Foundation,
http:$\sslash$www.blender.org). In virtual space, the selected cells or tissue fragments were viewed from an arbitrary direction, measured, hidden if necessary and manipulated as being of touchable
size. The automation of procedures mentioned is not known for any of the microscopy-oriented programs tested. Conventional 3D graphic environments are also incapable of this but allow cell lineages to
be traced and visualized manually with much more comfort. It should be noted that such a typical option for microscopy-oriented software as area and volume measurement is usually provided in
conventional 3D graphic software (Table 1).

\subsection*{Reconstruction of organelles in \textit{C. bracteata}}

Visualization of the architecture of organelles in the tuber phloem parenchyma cells of \textit{C. bracteata} shows the network organization of plastids (leucoplasts) as well as mitochondria in the
cell. Each plastid is connected to other plastids (possibly all) of the cell via tube-like structures with a diameter of about 500-600 nm (Fig. 8a). The length of these tubular connections ranges from
500 nm to 3 $\mu$m or more (Figs. 4a, 8). Their shape could represent a slightly undulating or curled curve (Fig. 8a). In the available data, neither branching of connections nor more than two
connections associated with one plastid was found. It could be preliminarily concluded that the tubular connections are the “traces” of plastid divisions which occurred without complete separation of
the daughter ones. However, more data are necessary to confirm this. These structures one could also refer to the stromules --- extremely variable and mobile protrusions of plastids (Hanson and
Sattararzadeh 2011).

The 3D reconstruction of the separate leucoplasts shows that the plastids have two basic shapes: i) convex with a crystal inside and ii) invaginated plastids filled with cytoplasm and even
mitochondria (Fig. 1a, 8a). Invaginated plastids prevailed in the sample of the cytoplasm reconstructed. Each of them has the shape of a sack with a quite narrow opening (about 100-200 nm in diameter)
which connects external and invaginated cytoplasm (arrows, Fig. 8b). This type of plastid corresponds to the plastolysome known to be the precursor of the autolytic vacuole in the process of
programmed cell death described in cells of \textit{Picea abies} L. embryos (Filonova et al. 2000).

The mitochondrial network was found to consist of a more or less uniformly branching tubular architecture with a transversal section diameter of about 300-400 nm (Fig. 8a). This network tightly fills
the space between each plastid and covers them almost completely. Thus, the cytoplasm of tuber phloem parenchyma cells of \textit{C. bracteata} represents a tightly packed volume filled with the
networks of the associated organelles. 

\end{multicols}

\begin{center}
\centerline{\includegraphics{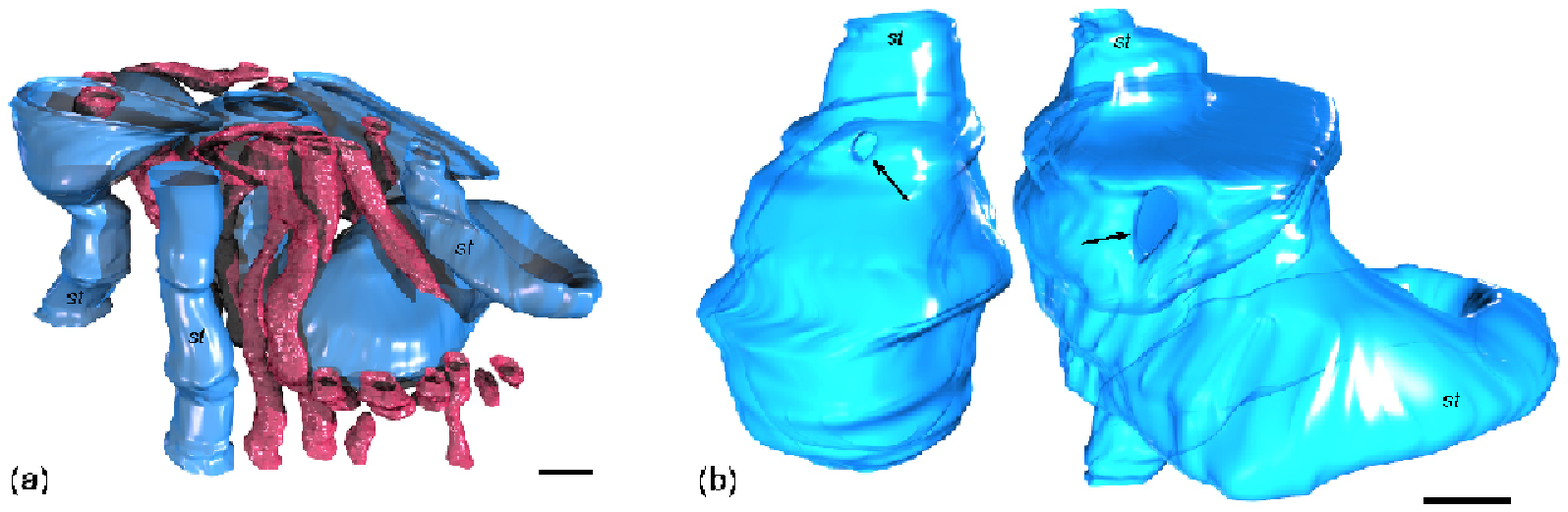}}
\captionof{figure}{\footnotesize Relations and topology of plastid and mitochondrial networks in tuber parenchyma cell of \textit{C. bracteata}. \textit{st} – stromule. Bar = 500 nm.\\
\textbf{(a)} – plastids and their tubular connections (blue). Branching network of mitochondria between plastids (red).\\
\textbf{(b)} – structure of invaginated plastids shown semitransparent. Internal space is connected via narrow openings (arrows).}
\label{Fig8}
\end{center}

\begin{multicols}{2}

\subsection*{Early embryogenesis of \textit{C. palustris}}

The reconstruction of the cellular architecture and the full genealogical tree of \textit{C. palustris} embryo were carried out according to our protocols (Fig. 4b, 9a). The embryo investigated
consisted of  $n = 17$ cells and was fixed at the blastomer stage when it has no internal cells. In the developmental history of this embryo, from 2 to 7 cell generations occurred in different cell
lineages (Fig. 9a). The zygote of \textit{C. palustris} was divided by a transversal plate into the apical and basal cells ($ca$ and $cb$, Fig. 9b). Due to the oblique division of the apical cell, two
unequal cells appeared; the smaller was wedge-shaped ($w$, Fig. 9c) while the bigger one became the initial cell of the quadrants ($iq$, Fig. 9c). This cell divided longitudinally to the apical-basal
axis of the embryo (Fig. 9d). As a result of the longitudinal divisions of each dyad cell, the quadrants were produced (Fig. 9e). All the quadrants of this embryo underwent divisions in different
planes, so the size of the octants was different and they did not manifest two precise tiers (Fig. 9f). The wedge-shaped cell underwent longitudinal division and then the two descendant cells divided
longitudinally and transversally respectively (Fig. 9g). As a result, the four descendant cells of the wedge-shaped cell also did not manifest a common tier between each other as well as any other
blastomer (Fig. 9h). The basal cell divided transversely and developed into the 2-celled suspensor (Fig. 9h).

\subsection*{Architecture and activity of the shoot apex of \textit{C.~palustris}}

\textit{C. palustris} has an alternative phyllotaxy. The reconstructed shoot apex represents a flattened convex structure with a slight prominence of $p0$ primordia and approximately half wrapped with
the first leaf primordia (Fig. 4c). The apex was composed of three relatively independent cell lineages whose cells were stretched along the orthostichy and divided the apex into three strips (shown
by color in Fig. 4c). One of them lies in a central position in relation to the orthostichy axis while the other two are lateral. Genealogical trees ($n = 208$ cells, up to 15 generations were traced)
are shown schematically in Fig. 10a-c. Due to the fragmentary nature of the tissue sample, it was impossible to relate three separate cells and three relatively independent lineages definitely to
their common ancestors. Thus, only $n-5 = 203$ cell divisions were found. Stem cells in a lineage were determined as cells whose adjacency to the neighboring cells repeats the adjacency of their own
merophytes in the course of all generations. Each cell lineage has its own one or two stem cells located at the surface (Fig. 4c, 8d, highlighted). All the subepidermal cells of the apex were found to
derive from the epidermal progenitors. These data indicate that the shoot apex of \textit{C. palustris} has an organization that differs from the “tunica-corpus” model.

\end{multicols}

\begin{figure}[!ht]
\centerline{\includegraphics{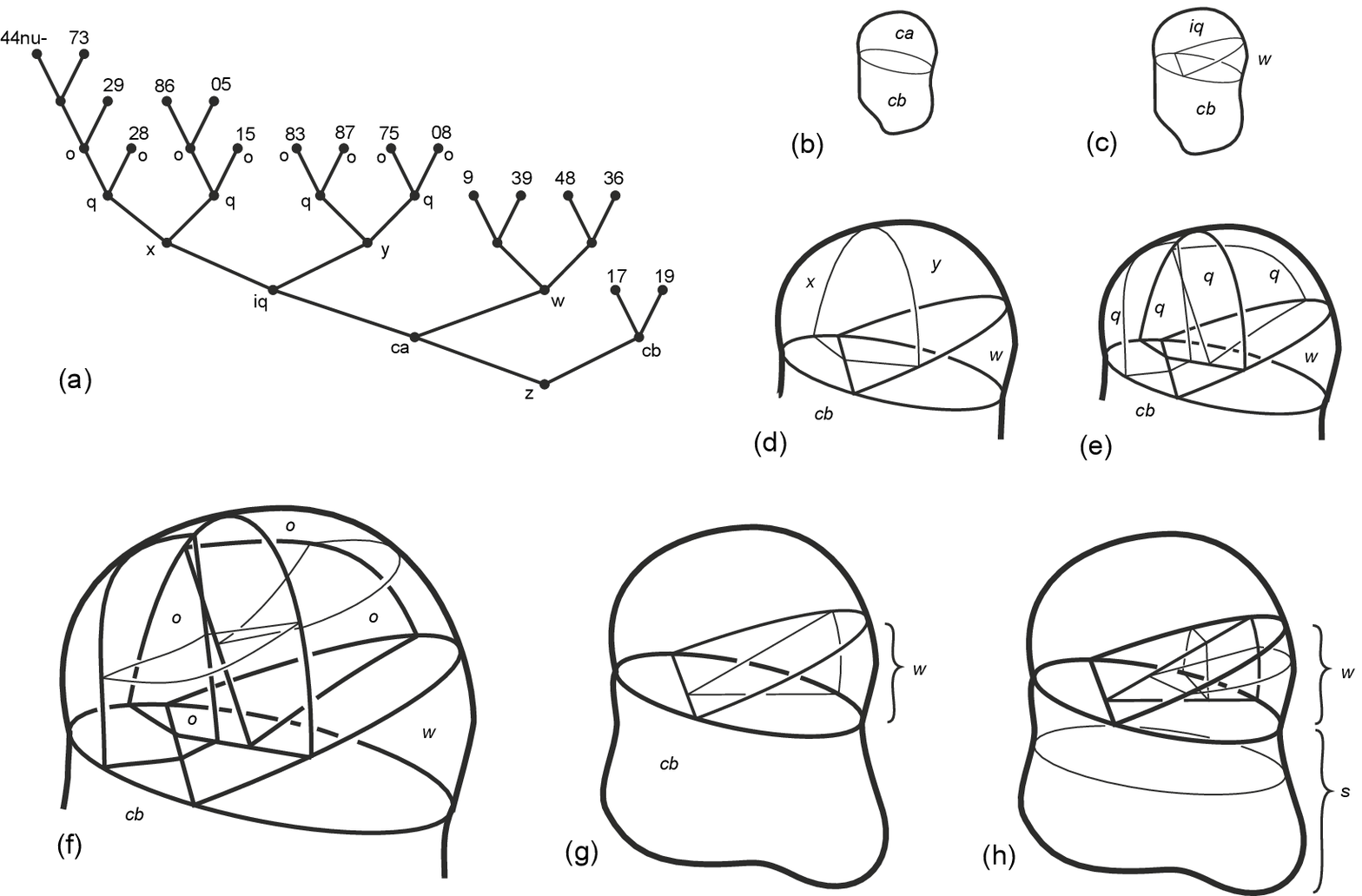}}
\caption{\footnotesize Developmental history of \textit{C. palustris embryo}. \textit{z} – zygote; \textit{ca} – apical cell; \textit{cb} – basal cell; \textit{w} – wedge-shaped cell; \textit{iq} –
quadrants` initial; \textit{x}, \textit{y} – precursors of quadrants (\textit{q}); \textit{o} – octants; \textit{s} – suspensor.\\
\textbf{(a)} – genealogical tree.\\
\textbf{(b-h)} – schemes of successive divisions in main cell lineages. In f, only 4 octants are shown.}
\label{Fig9}
\end{figure}

\begin{multicols}{2}

\section*{DISCUSSION}

\subsection*{Spatial architecture of organelles and tissues}

The spatial structure of organelles has been a key area of research for a long time. Since the introduction of transmission electron microscopy, many visualization methods have been designed, from
plastic models (Calvayrac and Lefort-Tran 1976; Gamalei and Pakhomova 1981; McFadden and Wetherbee 1982) to computer-based models giving the possibility of quantitative analysis (Mogensen et al. 1990)
and high resolution electron tomography (Frank et al. 2002). However, the 3-dimensional visualization of organelles is still not widely applied, perhaps due to the high cost of sophisticated hardware
and software. Thus, to date, there are few spatial data in modern botanical literature.

The existence of stromules is a notable example, where random TEM sections with high resolution did not help to perceive the whole spatial structure of the plastid networks inside a cell. Only
confocal microscopy with the optional spatial representation of structures, but with a lower resolution, has shown these structures in minute detail (Natesan et al. 2005). Our approach enables the
spatial information based on TEM to be seen, and our results suggest that the stromules could be a consequence of plastid division. Similar results on the division of amyloplasts in wheat endosperm
were observed by D.B. Bechtel and J.D. Wilson (2003). However, in general, as the stromules are a peculiar feature of plastids, they demonstrate a rather wide range of functions and ways of formation
(Hanson and Sattararzadeh 2011). Obviously, spatial information about the object is crucial for the resolution of many structural questions.

The first notable instrumental work in the 3D reconstruction of plant tissues was done by N. Hara in the “cell layer-by-cell layer” reconstruction of shoot apices of some dicots (Hara 1995). Confocal
microscopy and automatic 3D visualization by computer has enabled the quantitative analysis of cellular architecture (Gray et al. 1999). However, the optical limitations and demands for a large size
of cells have restricted the application of this approach, particularly for the description of formative tissues.

The protocols that we present in this paper could resolve the problems of spatial data acquisition and integration because they could be entirely carried out using open-source software installed on a
common computer since no considerable differences in results were found after using commercial and non-profit software. The device and software independence and flexibility of our protocols are based
on the application of the software which is most capable of performing an individual stage of the whole work. The application of a conventional 3D graphic editor even enables both types of source data
to be dealt with at the same time, especially in the case of the reconstruction of structures scanned at different scales with different types of microscope. The possession of detailed spatial
information about the object and the logic of developmental processes also opens up the possibility of revealing the object’s temporal changes during developmental processes, even without their
direct, real-time observation.

\subsection*{Spatiotemporal structure of plants}

The developmental history of the sample object has inevitably been coded in its spatial structure. Our protocols enable this history to be read from the subcellular organization and cell arrangement
in plant tissues. As mentioned above, the stromules appear to be traces of plastid division in tuber phloem parenchyma cells of \textit{C. bracteata}. Hence, the length, curvature and sequence of
connected plastids should reflect the genealogy and structure of plastid “lineages” inside a cell. The individual sequence and orientation of cell divisions inside cellular lineages of the tissue
sample could be compared with another sample in order to find the species-specific regularities of cellular architecture development.

The present protocol of cell lineage tracing using the geometrical properties of plant cells is a contemporary and rather technical extension of well known classical methods of plant developmental
biology. The absence of recent technologies in microscopy and computational devices at the beginning of the 20th century prevented the cell lineages of plant embryos to be traced back more than 3-5
generations (Sou\`eges 1937). Thus many important histological questions posed at that time are still unanswered. Our results unambiguously show the possibility of detecting the initial cells of the
lineage by tracing back almost all cell divisions (up to 15 generations at least) that occurred in the reconstructed embryo as well as in the shoot apex. The \textit{C. palustris} shoot apex activity
is in agreement with R. Korn’s point of view about the presence of apical cells undistinguishable from others in seed plant apices (Korn 1993). Our results also support the opinion about the presence
of periclinal divisions in the central zone of monocot shoot apex found earlier in maize and other grasses (Sharman 1940, 1943). However, they contradict the results of clonal analysis performed in
maize (Bossinger et al. 1992), where the absence of apical initials was concluded. We consider that the main feature of apical meristems is the presence of stem or autoreproductive cells whose
descendants self-maintain, participate in organogenesis via differentiation, and reproduce themselves via a set of often complex pathways as already shown by Barlow and colleagues for \textit{Psilotum
nudum} (Barlow et al. 2001).

\end{multicols}

\begin{center}
\centerline{\includegraphics{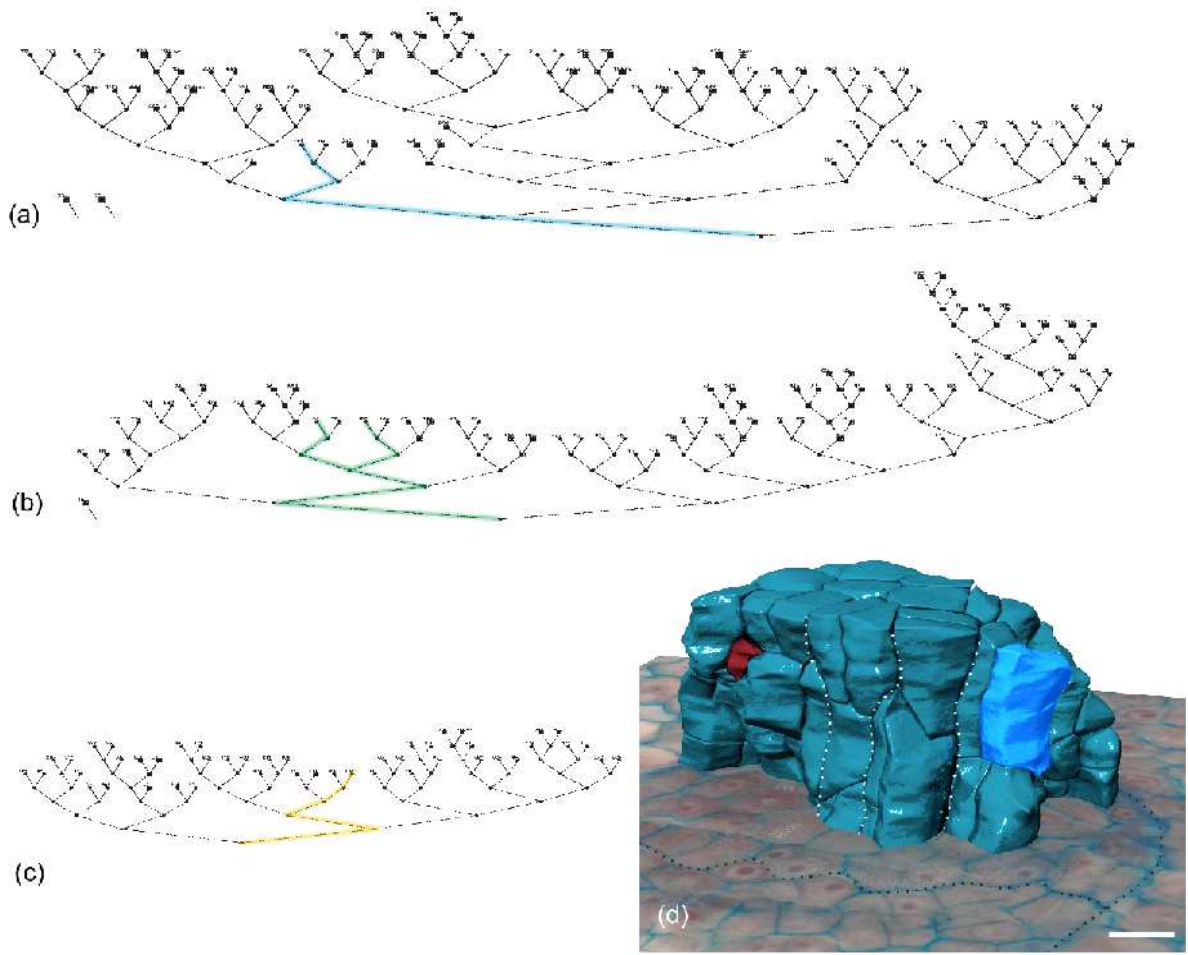}}
\captionof{figure}{\footnotesize Structure and developmental history of \textit{C. palustris} shoot apex.\\
\textbf{(a-c)} – genealogical trees of three main cell lineages. Lineages of stem cells are highlighted by color with respect to visualization in Fig. 4c. Internal cells denoted by squares.\\
\textbf{(d)} – microenvironment of stem cell (highlighted) of blue lineage. Cells of other lineages are removed. Merophytes (dotted white line) with periclinal divisions are clearly visible. Bar = 10
µm.}
\label{Fig10}
\end{center}

\begin{multicols}{2}

Current known approaches for cell lineage tracing provide only indirect or partial information. Classic clonal analysis deals with a cell lineage manifestation at the organ organization level. It is
based on the analysis of the distribution of cells with inherited marked features (plastid color, wax layer absence) of descendants in the developed organ (Bossinger et al. 1992; Dolan and Poethig
1998; Korn 2001; Harrison et al. 2007). Although conventional clonal analysis works well at the scale of the shoot system, our protocols help to reveal the formative tissues in detail. On the other
hand, plant cell geometry, namely the set edges and facets, being a natural “manuscript” of the developmental history of tissue, can be considerably altered or “rewritten” in the case of cell
separation or fusion, death or intercalar growth (Jarvis et al. 2003) like, for example, after aerenchima formation. Clonal analysis is insensitive to such tissue modifications. Undoubtedly, these two
approaches could well complement each other.

Automatic cell lineage tracing has only been done previously for the epiderm in the case of direct observation of the living shoot apex (Reddy et al. 2004), or in shoot apex replica analysis (Dumais
and Kwiatkowska 2001). The main advantage of both approaches is the ability to work with living and undamaged tissue but they are limited to the superficial tissues, exposed for observation, whilst
the shoot apex is uncovered by the developing leaves. This limitation was recently overcome by means of multi-angle imaging via confocal microscopy (Fernandez et al. 2010). The authors of this article
applied a repeated scanning of the object under different angles of observation, thus achieving impressive results in revealing real-time cell divisions in the whole volume of the \textit{Arabidopsis}
floral shoot apex. All these three quite powerful methodical approaches are artificially limited to just the “statistical” tracking of cell lineages since they record only the divisions occurring
during a period of real-time observation. For example, only up to 3 – 4 cell generations were recorded in most rapidly proliferating lineages (Reddy et al. 2004; Fernandez et al. 2010), which is
comparable to the classical works carried out almost a century ago (see Sou\`eges 1937). Our investigation was performed without high-end optical and computational equipment. Following just the logic
and geometrical properties of developmental processes, we have obtained rather significant results in lineage tracing. Our protocols are well adapted to confocal microscopy and conventional histology
and cytology, which deal with fixed material as well. Clearly, our approach can be used not instead of but as a useful addition to existing methods. The independence of the source data means that our
protocols can be applied even to paleobotanical material.

Genomic variability based on the known mutation rate in microsatellites is another potential way of automatically reconstructing cell lineage trees (Frumkin et al. 2005). Unfortunately, for the
moment, this method cannot reveal the genealogy of a single cell but only hundreds of cells which belong to a definite tissue. The revelation of cell lineages in living tissues with a fluorescent
marker in the nucleus is a very promising approach to date (Kurup et al. 2005), but it only works with a limited set of cell lineages in the organ, like a root, which has to be relatively transparent
and convenient for illumination and observation in a confocal microscope. Another significant technical achievement was made by P.J. Keller and colleagues using digital scanned laser light microscopy
in the reconstruction of the early embryogenesis of zebrafish (Keller et al. 2008). This sophisticated approach enables the genealogy and movements of hundreds of thousands of embryo cells to be
traced automatically using the fluorescent markers of their nuclei. We believe that a comparable result with plant embryos, which are not very transparent and enclosed in the seed tissues, could be
successfully achieved using our approach together with “virtual” microscopy. Hopefully, our protocol of cell lineage tracing could be automated and represented as a special software or environment,
which should assist in the manipulation of objects beyond human perception.

\section*{ACKNOWLEDGEMENTS}

This work was supported by the Program of Scientific Investigations of the Department of Biological Sciences of the Russian Academy of Sciences “Biological Resources of Russia” in
2009-2011. We thank two anonymous referees for all valuable comments and suggestions.

\bibliographystyle{plainnat}
\bibsep=3pt

\end{multicols}

\end{small}

\end{document}